\documentclass[cameraready]{Interspeech}
\title{Screening Matters: A Comparative Study of Conventional and\\Crowdsourced Listening Tests}

\author[orcid=0009-0008-4314-8730, equalcontribution]{Anika}{Treffehn}
\author[orcid=0000-0003-0939-7078, equalcontribution]{Andrea}{Eichenseer}
\author[orcid=0009-0009-6291-1786]{Emily}{Kratsch}
\author[orcid=0000-0003-0987-863X]{Nicola}{Pia}

\address{
    Fraunhofer-Institut f\"ur Integrierte Schaltungen IIS, Erlangen, Germany
}

\email{anika.treffehn@iis.fraunhofer.de}


\usepackage{graphicx, tabularx, enumitem, balance, flushend}
\usepackage[acronym]{glossaries}
\usepackage{pgfplots}
\pgfplotsset{compat=1.18}
\usepackage[utf8]{inputenc}
\usepackage{amsmath}
\usetikzlibrary{calc}  

\newacronym{AWS}{AWS}{Amazon Web Services}
\newacronym{AMR}{AMR}{Adaptive Multi-Rate}
\newacronym{AMR-WB}{AMR-WB}{Adaptive Multi-Rate -- Wideband}
\newacronym{DAC}{DAC}{Descript Audio Codec}
\newacronym{DCR}{DCR}{Degradation Category Rating}
\newacronym{DMOS}{DMOS}{Degradation Mean Opinion Score}
\newacronym{EVS}{EVS}{Enhanced Voice Services}
\newacronym{MJNDQ}{MJNDQ}{Modified Just-Noticeable Difference in Quality}
\newacronym{MNRU}{MNRU}{Modulated Noise Reference Unit}
\newacronym{MAE}{MAE}{Mean Absolute Error}
\newacronym{MTurk}{MTurk}{Amazon Mechanical Turk}
\newacronym{MUSHRA}{MUSHRA}{Multiple Stimuli with Hidden Reference and Anchor}
\newacronym{QoE}{QoE}{Quality of Experience}
\newacronym{RMSE}{RMSE}{Root Mean Square Error}
\newacronym{SD}{SD}{Standard Deviation}

\newcommand{\second}[1]{\SI{#1}{\second}}

\begin{document}

\maketitle

\begin{abstract}        
    Subjective evaluation remains the most reliable way of testing speech and audio coding techniques.
    Crowdsourcing the listening task is a cost-efficient and fast way of conducting this evaluation, but the quality of the results tends to be inferior to that of conventional listening tests done in the controlled environment of a laboratory.
    In this paper, classical and neural speech codecs are evaluated to compare P.808 against P.800 DCR tests.
    A statistical analysis is conducted to investigate the effectiveness of selected screening methods.
    The analysis shows that the crowdsourced evaluation can be improved by employing post-screening methods based on anchor ordering and rating span, and continuous screening methods like traps and gold standard questions, thus giving more value to the ratings obtained for the codecs under test. Based on these outcomes, a set of suitable screenings is proposed, for cost-effective, simplified, and bias-free enhancement of listening results.
\end{abstract}

\section{Introduction}

Evaluating the \gls{QoE}~\cite{qoe} of novel audio compression tools by means of subjective testing remains fundamental despite the plethora of available objective measures~\cite{objectivemeasures,objneural}.
With the success of neural speech and audio codecs, a need for robust measures of the output quality has arisen, as metrics developed for classical codecs fail in this scenario~\cite{crowdmushra}.
With subjective listening tests, human testers, 
who are also the end users of any speech or audio codec, 
provide valuable insight into the performance of the systems under test.
Recommendations such as P.800~\cite{P800} define a framework for subjective testing: they give instruction on the equipment, data, listener qualifications and more to ensure reliable results.
P.800 tests allow researchers to cover a larger set of audio data in their evaluation compared to other test methods like \gls{MUSHRA}~\cite{MUSHRA}.
This is especially useful for neural codecs as they might react less predictably to processing changes compared to classical codecs.
It is thus not enough to just cover a small set of audio items that are deemed critical (e.\,g., tonal, transient, high/low frequency content).
However, designing listening tests and conducting them with a sufficient amount of listeners is a very costly, time-consuming effort.

A more efficient means can be found in crowdsourced platforms such as \gls{MTurk} or Prolific\footnote{www.mturk.com, www.prolific.com, accessed February 2026.}.
Once set up--following, e.\,g., Recommendation P.808~\cite{P808}--the listening test can be conducted by any qualified worker on the platform.
This usually yields the desired number of results within only a couple of hours instead of weeks.
While results are obtained quickly, crowdsourced testing comes with certain drawbacks as there is no reliable way to effectively control the listening environment, the used equipment, or to provide assistance.
Consequently, the results are very often of poorer quality than those obtained by a controlled P.800 listening test.

In this paper, we evaluate a number of classical and neural speech and audio codecs to demonstrate how the quality of crowdsourced P.808 results declines compared against an equivalent P.800 test.
We analyze three categories of screening methods to detect invalid results: \textit{pre-screening} before the test, \textit{mid-screening} during the test, and \textit{post-screening} after the test.
Based on this analysis, we give a recommendation of methods to use for reliable improvement of crowdsourced test results.

\section{Listening experiments}
\label{sec:method}

\subsection{Test setup}
For the speech codec evaluation, two \gls{DCR}~\cite{P800} listening tests were conducted, employing the following \gls{DMOS} scale~\cite{MOS}:

\begin{enumerate}[leftmargin=3em, labelindent=2em, label=\arabic*:]
    \item Degradation is very annoying.
    \item Degradation is annoying.
    \item Degradation is slightly annoying.
    \item Degradation is audible but not annoying.    
    \item Degradation is inaudible.
\end{enumerate}

The first test follows Recommendation P.800 and was carried out in a controlled environment.
Listeners were selected based on their self-reported English language skill (at least B2 according to~\cite{CEFR}) and recent listening experience (no tests within the last two months).
Additionally, the hearing ability was checked by means of an audiogram and participants with significant hearing loss were excluded from the test.
The second test follows P.808 and was conducted via \gls{MTurk}.
Required qualifications were at least 80\% approval rate and 100 previously approved tasks, mild hearing loss at most, and no work experience in the field of audio/speech coding.
Sufficient English skill was ensured by self-reported English level and limiting worker location to Australia, Canada, Ireland, the UK, and the US.

\begin{table}[t]
  \caption{Listening test conditions with bit rate and bandwidth.}
  \label{tab:conditions}
  \centering
  \fontsize{8pt}{9pt}\selectfont 
  \begin{tabular}{ l l l l }
    \toprule
    \textbf{Index} &\textbf{Condition} &\textbf{kbit/s} &\textbf{kHz} \\
    \midrule
    c01& Reference           & -- & 16 \\
    c02--c04& MNRU~\cite{MNRU}, $Q=10,17,24$  & -- & 16 \\
    c05& 3.5 kHz low pass            & -- & 3.5 \\
    \midrule
    c06& Codec2~\cite{codec2}   & 2.4    & 4 \\
    c07, c08& AMR~\cite{amr}         & 4.75, 12.2   & 4 \\
    c09& AMR-WB~\cite{amr-wb}  & 12.65  & 8 \\
    c10, c11& EVS (WB, SWB)~\cite{evs}      & 13.2   & 8, 16 \\
    \midrule
    c12& FlowDec~\cite{flowdec} & 4.5 & 16 \\
    c13& Lyra~\cite{lyra}    & 3.2   & 8 \\
    c14, c15& DAC~\cite{dac}      & 1.5, 6.0     & 12 \\
    c16, c17& Mimi~\cite{mimi}    & 0.55, 1.1  & 12 \\
    c18 & SNAC~\cite{snac}    & 0.98  & 12 \\
    c19, c20& WavTokenizer~\cite{wavtokenizer} & 0.48, 0.9  & 12 \\
    \bottomrule
  \end{tabular}
\end{table}
\begin{table}[t]
  \caption{Distribution over demographic groups: female (f), male (m), age and distribution over listening panels.}
  \label{tab:demographic}
  \centering
  \fontsize{8pt}{9pt}\selectfont 
  \setlength{\tabcolsep}{5.15pt}
  \begin{tabularx}{\columnwidth}{ p{0.7cm}cccccc|cccccc|c}
    \toprule
    \textbf{Age} &\multicolumn{2}{c}{\textbf{15--29}} &\multicolumn{2}{c}{\textbf{30--50}} &\multicolumn{2}{c}{\textbf{$>$50}} & \multicolumn{6}{c}{\textbf{Listening panel}} &\\
    \textbf{Gender} & \textbf{f} & \textbf{m} & \textbf{f} & \textbf{m} & \textbf{f} & \textbf{m} & \textbf{1} & \textbf{2} & \textbf{3} & \textbf{4} & \textbf{5} & \textbf{6} & $\mathbf{\Sigma}$\\
    \midrule
    P.800 & 3 & 4 & 2 & 4 & 8 & 6 & 5 & 4 & 4 & 6 & 4 & 4 & 27\\
    P.808 & 5 & 5 & 6 & 7 & 5 & 5 & 4 & 5 & 6 & 7 & 6 & 5 & 33\\
   \bottomrule
  \end{tabularx}
\end{table}

The used proprietary test set comprises 24 monophonic English clean speech samples of \second{8} duration, distributed equally over two male and two female talkers, and each sample contains utterances of two distinct sentences and a pause in between.
Table~\ref{tab:conditions} lists all 20 conditions under test, totaling 480 stimuli divided into six panels according to~\cite{P808, Sup29}. Each participant listens to each condition exactly four times.
Including a training set to familiarize the listeners with the test method, the net duration amounts to approximately 32 minutes per participant.
Table~\ref{tab:demographic} gives the distribution of the listeners over demographic groups and listening panels.
Both listening test were created and conducted with a modified version of the webMUSHRA~\cite{webMUSHRA} framework.
Throughout the paper, error-based evaluations using \gls{MAE} and \gls{RMSE}~\cite{Robustness} are complemented by correlation-based evaluations using the Pearson correlation coefficient $r$ and Spearman's rank correlation coefficient $\rho$.
All measures are computed on mean ratings per condition.

\subsection{Listening results without screening}

\begin{figure*}[t]
    \centering
    \fontsize{8pt}{9pt}\selectfont 
    \begin{tikzpicture}[every node/.style={font=\footnotesize}]

\definecolor{darkgrey176}{RGB}{176,176,176}
\definecolor{grey}{RGB}{128,128,128}
\definecolor{blue1}{RGB}{34,94,168}
\definecolor{blue2}{RGB}{65,182,196}
\definecolor{blue3}{RGB}{161,158,180}
\definecolor{green1}{RGB}{0,104,55}
\definecolor{green2}{RGB}{49,163,84}
\definecolor{green3}{RGB}{120,198,121}
\definecolor{green4}{RGB}{194,230,153}
\definecolor{magenta1}{RGB}{122,1,119}
\definecolor{magenta2}{RGB}{197,27,138}
\definecolor{magenta3}{RGB}{215,181,216}
\definecolor{magenta4}{RGB}{215,48,31}
\definecolor{magenta5}{RGB}{252,141,89}
\definecolor{magenta6}{RGB}{253,204,138}

\begin{axis}[
tick align=outside,
tick pos=left,
x grid style={darkgrey176},
xmin=0.52, xmax=20.48,
xtick style={color=black},
xtick={1,2,3,4,5,6,7,8,9,10,11,12,13,14,15,16,17,18,19,20},
xticklabel style={rotate=90.0,anchor=east},
xticklabels={
  c01,
  c02,
  c03,
  c04,
  c05,
  c06,
  c07,
  c08,
  c09,
  c10,
  c11,
  c12,
  c13,
  c14,
  c15,
  c16,
  c17,
  c18,
  c19,
  c20
},
y grid style={darkgrey176},
ylabel={DMOS},
ymajorgrids,
ymin=0.8, ymax=5.45,
ytick style={color=black}
]
\addplot [very thin, grey, dashed]
table {
0.52 0.5
20.48 0.5
};
\addplot [very thin, grey, dashed]
table {
0.52 1.5
20.48 1.5
};
\addplot [very thin, grey, dashed]
table {
0.52 2.5
20.48 2.5
};
\addplot [very thin, grey, dashed]
table {
0.52 3.5
20.48 3.5
};
\addplot [very thin, grey, dashed]
table {
0.52 4.5
20.48 4.5
};

\draw[draw=none,fill=blue1] (axis cs:0.6,0) rectangle (axis cs:1.4,4.81481481481481);
\draw[draw=none,fill=blue2] (axis cs:1.6,0) rectangle (axis cs:2.4,1.32407407407407);
\draw[draw=none,fill=blue2] (axis cs:2.6,0) rectangle (axis cs:3.4,2.05555555555556);
\draw[draw=none,fill=blue2] (axis cs:3.6,0) rectangle (axis cs:4.4,3.31481481481481);
\draw[draw=none,fill=blue3] (axis cs:4.6,0) rectangle (axis cs:5.4,3.24074074074074);
\draw[draw=none,fill=green1] (axis cs:5.6,0) rectangle (axis cs:6.4,1.71296296296296);
\draw[draw=none,fill=green2] (axis cs:6.6,0) rectangle (axis cs:7.4,2.61111111111111);
\draw[draw=none,fill=green2] (axis cs:7.6,0) rectangle (axis cs:8.4,3.11111111111111);
\draw[draw=none,fill=green3] (axis cs:8.6,0) rectangle (axis cs:9.4,3.71296296296296);
\draw[draw=none,fill=green4] (axis cs:9.6,0) rectangle (axis cs:10.4,4.08333333333333);
\draw[draw=none,fill=green4] (axis cs:10.6,0) rectangle (axis cs:11.4,4.69444444444444);
\draw[draw=none,fill=magenta1] (axis cs:11.6,0) rectangle (axis cs:12.4,4.28703703703704);
\draw[draw=none,fill=magenta2] (axis cs:12.6,0) rectangle (axis cs:13.4,2.71296296296296);
\draw[draw=none,fill=magenta3] (axis cs:14.6,0) rectangle (axis cs:15.4,4.15740740740741);
\draw[draw=none,fill=magenta3] (axis cs:13.6,0) rectangle (axis cs:14.4,1.50925925925926);
\draw[draw=none,fill=magenta4] (axis cs:15.6,0) rectangle (axis cs:16.4,2.73148148148148);
\draw[draw=none,fill=magenta4] (axis cs:16.6,0) rectangle (axis cs:17.4,3.78703703703704);
\draw[draw=none,fill=magenta5] (axis cs:17.6,0) rectangle (axis cs:18.4,4.41666666666667);
\draw[draw=none,fill=magenta6] (axis cs:18.6,0) rectangle (axis cs:19.4,3.78703703703704);
\draw[draw=none,fill=magenta6] (axis cs:19.6,0) rectangle (axis cs:20.4,4.47222222222222);
\path [draw=black, semithick]
(axis cs:1,4.73682571686683)
--(axis cs:1,4.8928039127628);

\path [draw=black, semithick]
(axis cs:2,1.19820544235002)
--(axis cs:2,1.44994270579813);

\path [draw=black, semithick]
(axis cs:3,1.899065114616)
--(axis cs:3,2.21204599649511);

\path [draw=black, semithick]
(axis cs:4,3.14060248672377)
--(axis cs:4,3.48902714290586);

\path [draw=black, semithick]
(axis cs:5,3.07389285649029)
--(axis cs:5,3.4075886249912);

\path [draw=black, semithick]
(axis cs:6,1.5518305278233)
--(axis cs:6,1.87409539810263);

\path [draw=black, semithick]
(axis cs:7,2.43092326763605)
--(axis cs:7,2.79129895458618);

\path [draw=black, semithick]
(axis cs:8,2.93942605461327)
--(axis cs:8,3.28279616760896);

\path [draw=black, semithick]
(axis cs:9,3.53989401118533)
--(axis cs:9,3.88603191474059);

\path [draw=black, semithick]
(axis cs:10,3.92204814789498)
--(axis cs:10,4.24461851877168);

\path [draw=black, semithick]
(axis cs:11,4.59639922584704)
--(axis cs:11,4.79248966304184);

\path [draw=black, semithick]
(axis cs:12,4.14335430402127)
--(axis cs:12,4.43071977005281);

\path [draw=black, semithick]
(axis cs:13,2.52693358565747)
--(axis cs:13,2.89899234026846);

\path [draw=black, semithick]
(axis cs:15,3.9899794051393)
--(axis cs:15,4.32483540967552);

\path [draw=black, semithick]
(axis cs:14,1.38164203102309)
--(axis cs:14,1.63687648749543);

\path [draw=black, semithick]
(axis cs:16,2.544461964774)
--(axis cs:16,2.91850099818897);

\path [draw=black, semithick]
(axis cs:17,3.64568690074073)
--(axis cs:17,3.92838717333335);

\path [draw=black, semithick]
(axis cs:18,4.27281265332423)
--(axis cs:18,4.5605206800091);

\path [draw=black, semithick]
(axis cs:19,3.60644819736806)
--(axis cs:19,3.96762587670601);

\path [draw=black, semithick]
(axis cs:20,4.34212072728594)
--(axis cs:20,4.6023237171585);

\draw (axis cs:1,5.2) ++(0pt,10pt) node[
  scale=1.0,
  anchor=east,
  text=black,
  rotate=90.0
]{4.81};
\draw (axis cs:2,5.2) ++(0pt,10pt) node[
  scale=1.0,
  anchor=east,
  text=black,
  rotate=90.0
]{1.32};
\draw (axis cs:3,5.2) ++(0pt,10pt) node[
  scale=1.0,
  anchor=east,
  text=black,
  rotate=90.0
]{2.06};
\draw (axis cs:4,5.2) ++(0pt,10pt) node[
  scale=1.0,
  anchor=east,
  text=black,
  rotate=90.0
]{3.31};
\draw (axis cs:5,5.2) ++(0pt,10pt) node[
  scale=1.0,
  anchor=east,
  text=black,
  rotate=90.0
]{3.24};
\draw (axis cs:6,5.2) ++(0pt,10pt) node[
  scale=1.0,
  anchor=east,
  text=black,
  rotate=90.0
]{1.71};
\draw (axis cs:7,5.2) ++(0pt,10pt) node[
  scale=1.0,
  anchor=east,
  text=black,
  rotate=90.0
]{2.61};
\draw (axis cs:8,5.2) ++(0pt,10pt) node[
  scale=1.0,
  anchor=east,
  text=black,
  rotate=90.0
]{3.11};
\draw (axis cs:9,5.2) ++(0pt,10pt) node[
  scale=1.0,
  anchor=east,
  text=black,
  rotate=90.0
]{3.71};
\draw (axis cs:10,5.2) ++(0pt,10pt) node[
  scale=1.0,
  anchor=east,
  text=black,
  rotate=90.0
]{4.08};
\draw (axis cs:11,5.2) ++(0pt,10pt) node[
  scale=1.0,
  anchor=east,
  text=black,
  rotate=90.0
]{4.69};
\draw (axis cs:12,5.2) ++(0pt,10pt) node[
  scale=1.0,
  anchor=east,
  text=black,
  rotate=90.0
]{4.29};
\draw (axis cs:13,5.2) ++(0pt,10pt) node[
  scale=1.0,
  anchor=east,
  text=black,
  rotate=90.0
]{2.71};
\draw (axis cs:14,5.2) ++(0pt,10pt) node[
  scale=1.0,
  anchor=east,
  text=black,
  rotate=90.0
]{4.16};
\draw (axis cs:15,5.2) ++(0pt,10pt) node[
  scale=1.0,
  anchor=east,
  text=black,
  rotate=90.0
]{1.51};
\draw (axis cs:16,5.2) ++(0pt,10pt) node[
  scale=1.0,
  anchor=east,
  text=black,
  rotate=90.0
]{2.73};
\draw (axis cs:17,5.2) ++(0pt,10pt) node[
  scale=1.0,
  anchor=east,
  text=black,
  rotate=90.0
]{3.79};
\draw (axis cs:18,5.2) ++(0pt,10pt) node[
  scale=1.0,
  anchor=east,
  text=black,
  rotate=90.0
]{4.42};
\draw (axis cs:19,5.2) ++(0pt,10pt) node[
  scale=1.0,
  anchor=east,
  text=black,
  rotate=90.0
]{3.79};
\draw (axis cs:20,5.2) ++(0pt,10pt) node[
  scale=1.0,
  anchor=east,
  text=black,
  rotate=90.0
]{4.47};
\end{axis}

\end{tikzpicture}
    \hspace{0.9cm}
    \begin{tikzpicture}[every node/.style={font=\footnotesize}]

\definecolor{darkgrey176}{RGB}{176,176,176}
\definecolor{grey}{RGB}{128,128,128}
\definecolor{blue1}{RGB}{34,94,168}
\definecolor{blue2}{RGB}{65,182,196}
\definecolor{blue3}{RGB}{161,158,180}
\definecolor{green1}{RGB}{0,104,55}
\definecolor{green2}{RGB}{49,163,84}
\definecolor{green3}{RGB}{120,198,121}
\definecolor{green4}{RGB}{194,230,153}
\definecolor{magenta1}{RGB}{122,1,119}
\definecolor{magenta2}{RGB}{197,27,138}
\definecolor{magenta3}{RGB}{215,181,216}
\definecolor{magenta4}{RGB}{215,48,31}
\definecolor{magenta5}{RGB}{252,141,89}
\definecolor{magenta6}{RGB}{253,204,138}

\begin{axis}[
tick align=outside,
tick pos=left,
x grid style={darkgrey176},
xmin=0.52, xmax=20.48,
xtick style={color=black},
xtick={1,2,3,4,5,6,7,8,9,10,11,12,13,14,15,16,17,18,19,20},
xticklabel style={rotate=90.0,anchor=east},
xticklabels={
  c01,
  c02,
  c03,
  c04,
  c05,
  c06,
  c07,
  c08,
  c09,
  c10,
  c11,
  c12,
  c13,
  c14,
  c15,
  c16,
  c17,
  c18,
  c19,
  c20,
},
y grid style={darkgrey176},
ylabel={DMOS},
ymajorgrids,
ymin=0.8, ymax=5.45,
ytick style={color=black}
]
\addplot [very thin, grey, dashed]
table {
0.52 0.5
20.48 0.5
};
\addplot [very thin, grey, dashed]
table {
0.52 1.5
20.48 1.5
};
\addplot [very thin, grey, dashed]
table {
0.52 2.5
20.48 2.5
};
\addplot [very thin, grey, dashed]
table {
0.52 3.5
20.48 3.5
};
\addplot [very thin, grey, dashed]
table {
0.52 4.5
20.48 4.5
};

\draw[draw=none,fill=blue1] (axis cs:0.6,0) rectangle (axis cs:1.4,3.70454545454545);
\draw[draw=none,fill=blue2] (axis cs:1.6,0) rectangle (axis cs:2.4,2.1969696969697);
\draw[draw=none,fill=blue2] (axis cs:2.6,0) rectangle (axis cs:3.4,3.12121212121212);
\draw[draw=none,fill=blue2] (axis cs:3.6,0) rectangle (axis cs:4.4,3.57575757575758);
\draw[draw=none,fill=blue3] (axis cs:4.6,0) rectangle (axis cs:5.4,3.12121212121212);
\draw[draw=none,fill=green1] (axis cs:5.6,0) rectangle (axis cs:6.4,2.5);
\draw[draw=none,fill=green2] (axis cs:6.6,0) rectangle (axis cs:7.4,2.93939393939394);
\draw[draw=none,fill=green2] (axis cs:7.6,0) rectangle (axis cs:8.4,3.03787878787879);
\draw[draw=none,fill=green3] (axis cs:8.6,0) rectangle (axis cs:9.4,3.48484848484848);
\draw[draw=none,fill=green4] (axis cs:9.6,0) rectangle (axis cs:10.4,3.35606060606061);
\draw[draw=none,fill=green4] (axis cs:10.6,0) rectangle (axis cs:11.4,3.81818181818182);
\draw[draw=none,fill=magenta1] (axis cs:11.6,0) rectangle (axis cs:12.4,3.67424242424242);
\draw[draw=none,fill=magenta2] (axis cs:12.6,0) rectangle (axis cs:13.4,3.18181818181818);
\draw[draw=none,fill=magenta3] (axis cs:14.6,0) rectangle (axis cs:15.4,3.6969696969697);
\draw[draw=none,fill=magenta3] (axis cs:13.6,0) rectangle (axis cs:14.4,2.62121212121212);
\draw[draw=none,fill=magenta4] (axis cs:15.6,0) rectangle (axis cs:16.4,3.14393939393939);
\draw[draw=none,fill=magenta4] (axis cs:16.6,0) rectangle (axis cs:17.4,3.65151515151515);
\draw[draw=none,fill=magenta5] (axis cs:17.6,0) rectangle (axis cs:18.4,3.76515151515152);
\draw[draw=none,fill=magenta6] (axis cs:18.6,0) rectangle (axis cs:19.4,3.50757575757576);
\draw[draw=none,fill=magenta6] (axis cs:19.6,0) rectangle (axis cs:20.4,3.59848484848485);
\path [draw=black, semithick]
(axis cs:1,3.51165470622358)
--(axis cs:1,3.89743620286733);

\path [draw=black, semithick]
(axis cs:2,1.97645613960962)
--(axis cs:2,2.41748325432977);

\path [draw=black, semithick]
(axis cs:3,2.93146631436893)
--(axis cs:3,3.31095792805531);

\path [draw=black, semithick]
(axis cs:4,3.40306313747764)
--(axis cs:4,3.74845201403751);

\path [draw=black, semithick]
(axis cs:5,2.91790121620573)
--(axis cs:5,3.32452302621852);

\path [draw=black, semithick]
(axis cs:6,2.28868443533536)
--(axis cs:6,2.71131556466464);

\path [draw=black, semithick]
(axis cs:7,2.74192865066366)
--(axis cs:7,3.13685922812422);

\path [draw=black, semithick]
(axis cs:8,2.84307792568771)
--(axis cs:8,3.23267965006987);

\path [draw=black, semithick]
(axis cs:9,3.30913968467368)
--(axis cs:9,3.66055728502329);

\path [draw=black, semithick]
(axis cs:10,3.1795347401488)
--(axis cs:10,3.53258647197241);

\path [draw=black, semithick]
(axis cs:11,3.65515274485431)
--(axis cs:11,3.98121089150933);

\path [draw=black, semithick]
(axis cs:12,3.51440882739235)
--(axis cs:12,3.8340760210925);

\path [draw=black, semithick]
(axis cs:13,3.00318355015352)
--(axis cs:13,3.36045281348285);

\path [draw=black, semithick]
(axis cs:15,3.51943134151815)
--(axis cs:15,3.87450805242124);

\path [draw=black, semithick]
(axis cs:14,2.3986020047504)
--(axis cs:14,2.84382223767384);

\path [draw=black, semithick]
(axis cs:16,2.96367751622838)
--(axis cs:16,3.3242012716504);

\path [draw=black, semithick]
(axis cs:17,3.47143542968196)
--(axis cs:17,3.83159487334835);

\path [draw=black, semithick]
(axis cs:18,3.59936359558335)
--(axis cs:18,3.93093943471968);

\path [draw=black, semithick]
(axis cs:19,3.32686627842402)
--(axis cs:19,3.68828523672749);

\path [draw=black, semithick]
(axis cs:20,3.41011443256277)
--(axis cs:20,3.78685526440692);

\draw (axis cs:1,5.2) ++(0pt,10pt) node[
  scale=1.0,
  anchor=east,
  text=black,
  rotate=90.0
]{3.70};
\draw (axis cs:2,5.2) ++(0pt,10pt) node[
  scale=1.0,
  anchor=east,
  text=black,
  rotate=90.0
]{2.20};
\draw (axis cs:3,5.2) ++(0pt,10pt) node[
  scale=1.0,
  anchor=east,
  text=black,
  rotate=90.0
]{3.12};
\draw (axis cs:4,5.2) ++(0pt,10pt) node[
  scale=1.0,
  anchor=east,
  text=black,
  rotate=90.0
]{3.58};
\draw (axis cs:5,5.2) ++(0pt,10pt) node[
  scale=1.0,
  anchor=east,
  text=black,
  rotate=90.0
]{3.12};
\draw (axis cs:6,5.2) ++(0pt,10pt) node[
  scale=1.0,
  anchor=east,
  text=black,
  rotate=90.0
]{2.50};
\draw (axis cs:7,5.2) ++(0pt,10pt) node[
  scale=1.0,
  anchor=east,
  text=black,
  rotate=90.0
]{2.94};
\draw (axis cs:8,5.2) ++(0pt,10pt) node[
  scale=1.0,
  anchor=east,
  text=black,
  rotate=90.0
]{3.04};
\draw (axis cs:9,5.2) ++(0pt,10pt) node[
  scale=1.0,
  anchor=east,
  text=black,
  rotate=90.0
]{3.48};
\draw (axis cs:10,5.2) ++(0pt,10pt) node[
  scale=1.0,
  anchor=east,
  text=black,
  rotate=90.0
]{3.36};
\draw (axis cs:11,5.2) ++(0pt,10pt) node[
  scale=1.0,
  anchor=east,
  text=black,
  rotate=90.0
]{3.82};
\draw (axis cs:12,5.2) ++(0pt,10pt) node[
  scale=1.0,
  anchor=east,
  text=black,
  rotate=90.0
]{3.67};
\draw (axis cs:13,5.2) ++(0pt,10pt) node[
  scale=1.0,
  anchor=east,
  text=black,
  rotate=90.0
]{3.18};
\draw (axis cs:14,5.2) ++(0pt,10pt) node[
  scale=1.0,
  anchor=east,
  text=black,
  rotate=90.0
]{3.70};
\draw (axis cs:15,5.2) ++(0pt,10pt) node[
  scale=1.0,
  anchor=east,
  text=black,
  rotate=90.0
]{2.62};
\draw (axis cs:16,5.2) ++(0pt,10pt) node[
  scale=1.0,
  anchor=east,
  text=black,
  rotate=90.0
]{3.14};
\draw (axis cs:17,5.2) ++(0pt,10pt) node[
  scale=1.0,
  anchor=east,
  text=black,
  rotate=90.0
]{3.65};
\draw (axis cs:18,5.2) ++(0pt,10pt) node[
  scale=1.0,
  anchor=east,
  text=black,
  rotate=90.0
]{3.77};
\draw (axis cs:19,5.2) ++(0pt,10pt) node[
  scale=1.0,
  anchor=east,
  text=black,
  rotate=90.0
]{3.51};
\draw (axis cs:20,5.2) ++(0pt,10pt) node[
  scale=1.0,
  anchor=east,
  text=black,
  rotate=90.0
]{3.60};
\end{axis}

\end{tikzpicture}
    \caption{P.800 (left) and P.808 (right) results without any screening applied.}
    \label{fig:comparison}
\end{figure*}
\begin{table}[t]
  \caption{DMOS distribution in \%.}
  \label{tab:scoredist}
  \centering
  \fontsize{8pt}{9pt}\selectfont 
  \begin{tabularx}{0.9\columnwidth}{ Xcccccc}
    \toprule
    \textbf{Score} & \textbf{1} & \textbf{2} & \textbf{3} & \textbf{4} & \textbf{5} & \textbf{Mean}\\
    \midrule
    P.800 & 12.3 & 16.2 & 23.0 & 23.6 & 25.0 & 3.32\\
    P.808 & 8.6 & 18.6 & 23.8 & 33.9 & 15.2 & 3.28\\
    \bottomrule
  \end{tabularx}
\end{table}

Figure~\ref{fig:comparison} shows the results per condition obtained for both listening tests.
It is immediately visible that the relative ordering of the conditions is very similar, but the effectively used rating span is significantly decreased in the P.808 test.
The correlation coefficients $r=0.929$ and $\rho=0.929$ of the DMOS values show that the two tests mostly agree on which conditions perform better or worse, but the \gls{MAE} of $0.573$ and the \gls{RMSE} of $0.659$ between the two tests shows that, on average, the rating differs by more than half a DMOS point.
The differences even amplify for conditions ranked very high and very low. The rating of the reference condition c01 decreases by $1.11$, 
while the rating of the worst anchor condition c02 increases by $0.88$ compared to the P.800 test.

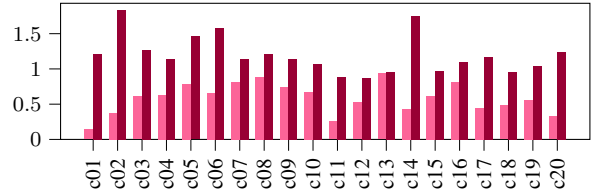
\begin{figure}[t]
    \centering
    \fontsize{8pt}{9pt}\selectfont 
    \begin{tikzpicture}

\definecolor{darkgrey176}{RGB}{176,176,176}
\definecolor{darkorange25512714}{RGB}{152,0,52}
\definecolor{lightgrey204}{RGB}{204,204,204}
\definecolor{steelblue31119180}{RGB}{252,100,152}

\begin{axis}[
tick align=outside,
tick pos=left,
x grid style={darkgrey176},
xmin=-1.335, xmax=20.335,
xtick style={color=black},
xtick={0,1,2,3,4,5,6,7,8,9,10,11,12,13,14,15,16,17,18,19},
xticklabel style={rotate=90.0,anchor=east},
xticklabels={c01,c02,c03,c04,c05,c06,c07,c08,c09,c10,c11,c12,c13,c14,c15,c16,c17,c18,c19,c20},
y grid style={darkgrey176},
ymin=0, ymax=1.93104166666667,
ytick style={color=black},
scale only axis,
width=7.0cm,
height=1.8cm, 
]
\draw[draw=none,fill=steelblue31119180] (axis cs:-0.35,0) rectangle (axis cs:0,0.140277777777778);

\draw[draw=none,fill=steelblue31119180] (axis cs:0.65,0) rectangle (axis cs:1,0.377777777777778);
\draw[draw=none,fill=steelblue31119180] (axis cs:1.65,0) rectangle (axis cs:2,0.607638888888889);
\draw[draw=none,fill=steelblue31119180] (axis cs:2.65,0) rectangle (axis cs:3,0.631944444444444);
\draw[draw=none,fill=steelblue31119180] (axis cs:3.65,0) rectangle (axis cs:4,0.789583333333333);
\draw[draw=none,fill=steelblue31119180] (axis cs:4.65,0) rectangle (axis cs:5,0.656944444444444);
\draw[draw=none,fill=steelblue31119180] (axis cs:5.65,0) rectangle (axis cs:6,0.820138888888889);
\draw[draw=none,fill=steelblue31119180] (axis cs:6.65,0) rectangle (axis cs:7,0.888194444444444);
\draw[draw=none,fill=steelblue31119180] (axis cs:7.65,0) rectangle (axis cs:8,0.739583333333333);
\draw[draw=none,fill=steelblue31119180] (axis cs:8.65,0) rectangle (axis cs:9,0.675694444444444);
\draw[draw=none,fill=steelblue31119180] (axis cs:9.65,0) rectangle (axis cs:10,0.254166666666667);
\draw[draw=none,fill=steelblue31119180] (axis cs:10.65,0) rectangle (axis cs:11,0.523611111111111);
\draw[draw=none,fill=steelblue31119180] (axis cs:11.65,0) rectangle (axis cs:12,0.935416666666667);
\draw[draw=none,fill=steelblue31119180] (axis cs:13.65,0) rectangle (axis cs:14,0.61875);
\draw[draw=none,fill=steelblue31119180] (axis cs:12.65,0) rectangle (axis cs:13,0.43125);
\draw[draw=none,fill=steelblue31119180] (axis cs:14.65,0) rectangle (axis cs:15,0.816666666666667);
\draw[draw=none,fill=steelblue31119180] (axis cs:15.65,0) rectangle (axis cs:16,0.445833333333333);
\draw[draw=none,fill=steelblue31119180] (axis cs:16.65,0) rectangle (axis cs:17,0.480555555555556);
\draw[draw=none,fill=steelblue31119180] (axis cs:17.65,0) rectangle (axis cs:18,0.556944444444444);
\draw[draw=none,fill=steelblue31119180] (axis cs:18.65,0) rectangle (axis cs:19,0.325694444444444);
\draw[draw=none,fill=darkorange25512714] (axis cs:2.77555756156289e-17,0) rectangle (axis cs:0.35,1.21597222222222);

\draw[draw=none,fill=darkorange25512714] (axis cs:1,0) rectangle (axis cs:1.35,1.8390873015873);
\draw[draw=none,fill=darkorange25512714] (axis cs:2,0) rectangle (axis cs:2.35,1.26636904761905);
\draw[draw=none,fill=darkorange25512714] (axis cs:3,0) rectangle (axis cs:3.35,1.14166666666667);
\draw[draw=none,fill=darkorange25512714] (axis cs:4,0) rectangle (axis cs:4.35,1.45902777777778);
\draw[draw=none,fill=darkorange25512714] (axis cs:5,0) rectangle (axis cs:5.35,1.58075396825397);
\draw[draw=none,fill=darkorange25512714] (axis cs:6,0) rectangle (axis cs:6.35,1.14335317460317);
\draw[draw=none,fill=darkorange25512714] (axis cs:7,0) rectangle (axis cs:7.35,1.21130952380952);
\draw[draw=none,fill=darkorange25512714] (axis cs:8,0) rectangle (axis cs:8.35,1.13521825396825);
\draw[draw=none,fill=darkorange25512714] (axis cs:9,0) rectangle (axis cs:9.35,1.06517857142857);
\draw[draw=none,fill=darkorange25512714] (axis cs:10,0) rectangle (axis cs:10.35,0.887400793650794);
\draw[draw=none,fill=darkorange25512714] (axis cs:11,0) rectangle (axis cs:11.35,0.872519841269841);
\draw[draw=none,fill=darkorange25512714] (axis cs:12,0) rectangle (axis cs:12.35,0.954861111111111);
\draw[draw=none,fill=darkorange25512714] (axis cs:14,0) rectangle (axis cs:14.35,0.973809523809524);
\draw[draw=none,fill=darkorange25512714] (axis cs:13,0) rectangle (axis cs:13.35,1.74821428571429);
\draw[draw=none,fill=darkorange25512714] (axis cs:15,0) rectangle (axis cs:15.35,1.09295634920635);
\draw[draw=none,fill=darkorange25512714] (axis cs:16,0) rectangle (axis cs:16.35,1.16507936507937);
\draw[draw=none,fill=darkorange25512714] (axis cs:17,0) rectangle (axis cs:17.35,0.956150793650794);
\draw[draw=none,fill=darkorange25512714] (axis cs:18,0) rectangle (axis cs:18.35,1.0390873015873);
\draw[draw=none,fill=darkorange25512714] (axis cs:19,0) rectangle (axis cs:19.35,1.24454365079365);
\end{axis}

\end{tikzpicture}
    \caption{Mean variance; light: P.800, dark: P.808.}
    \label{fig:variance}
\end{figure}

Further investigation yields two key aspects that cause these differences.
\textbf{Unused scale}: Table~\ref{tab:scoredist} shows that the P.808 participants tend towards rating conditions moderately and use the extreme ratings (1 and 5) less often compared to the participants of the P.800 test.
\textbf{Increased variance}: As visible in Figure~\ref{fig:variance}, the variances within conditions are significantly increased for P.808 results for most conditions.
This is especially apparent for the conditions with high and poor ratings. The increased variance pushes the condition mean towards the center of the rating scale and away from the extremes due to the finite scale.

\subsection{Screening methods}
\label{sec:screening_method}
To alleviate the discovered weaknesses, we employ three categories of screening methods on the P.808 results.

\subsubsection{Pre-screening}

Listener (dis)qualification before the test is ideal as it saves time and money and does not alter the distribution over demographic groups and listening panels in an unfavorable way.

\textbf{Pretest}:
A pretest similar to the \gls{MJNDQ} test~\cite{P808, jnd} and adapted for the \gls{DCR} methodology is employed to check the listener's hardware, environment, and hearing and listening differentiation ability.
Our pretest contains ten items for which the listeners must identify the condition A or B that is qualitatively more similar to the given reference.
Degradations employed include bandwidth reduction, added noise, and compression artifacts to cover typically encountered aberrations.

\textbf{Questionnaire}:
Questions as recommended in~\cite{P808} must be answered favorably: ``Have you participated in a subjective test (opinion test) in the last seven days?'', ``Have you participated in an audio listening test in the last two weeks?'', ``Is your current environment quiet and free from disturbances? If not, are you able to move to a more suitable location?'', ``This test requires the use of headphones. Are you currently wearing wired headphones, or do you have access to them?''
The goal is to make the listener aware of the hardware and environment requirements and to remove non-naive workers with too much experience.

\subsubsection{Mid-screening}

Monitoring the test is an efficient means to remove listeners as soon as their lack of qualification becomes obvious.
Direct feedback can be given as to the reason for failure.

\textbf{Traps}:
The main listening test contains randomly placed traps~\cite{P808, naderi15_interspeech} that require the listener to click a rating as demanded by the corresponding audio.
The test can be stopped immediately if a participant exceeds a predefined number of failures.
This mechanism checks for basic language understanding and, purely by existing, increases attention.

\textbf{Gold standard questions}:
P.808 recommends the use of \textit{gold standard} questions based on reference and anchor ratings.
We only make use of the reference rating: Users are screened by observing their lowest reference rating and comparing it to a threshold.
The goal is again to test the worker's attention and the ability to detect perfect quality within the used setup.
From experience, anchor ratings tend to be very subjective, and are thus less reliable for screening.

\subsubsection{Post-screening}
In \cite{P808}, screening for unexpected patterns and outliers is recommended, but not many details are given.
Post-screening can be very effective, but it does not exhibit the advantages of the previous screening methods.

\textbf{Rating span}:
To ensure a reasonable spread across the scale, the rating span is computed as the difference between the listener's mean ratings of the reference and the lowest anchor condition (here c02). 
Care must be taken to ascertain c02 is in fact the condition with the strongest degradation in the listening test. 
The applied screening makes sure that all participants fulfill or exceed a defined threshold for the span.

\textbf{Anchor ordering}:
The ordering of the \gls{MNRU} anchors (c02 $\leq$ c03 $\leq$ c04 $\leq$ c01) is used as a screening method to ensure the listener is able to accurately judge different grades of degradation~\cite{Sup29}.
The screening is based on the number of correctly fulfilled inequalities for each participant's average ratings (from 0 to 3).
For this screening to work, it is necessary that the \gls{MNRU} conditions are designed to sufficiently cover the DMOS scale by adjusting the noise power level controlled by $Q$.

For both of the presented post-screening methods, it is important to consider the number of times a condition is rated per user.
If there are just one or two ratings per condition, the mean values across one condition will become less meaningful and small outliers can lead to different screening outcomes.

\section{Statistical analysis and evaluation}
\label{sec:results}

The effects of the proposed screening methods on the P.808 listening test results are analyzed and the results
evaluated by their \textit{closeness} to the P.800 results with the assumption that the P.800 results match the desired \gls{QoE}.

\subsection{Analysis of pre-screening methods}
\textbf{Pretest}:
The majority of participants performed well on the pretest: 24 of the participants attained at least nine out of ten correct answers. 
It was observed that the pretest correctness is a very weak indicator of result quality as the best results are gained by setting the threshold to five out of ten correct answers, yielding $r=0.941$, $\rho=0.933$, $\mathrm{RMSE}=0.654$, and $\mathrm{MAE=0.561}$ with 31 participants, which is just marginally better than the baseline without screening. 
With stricter screening, all measures decrease, and even fall below the baseline values.
We conclude that the pretest in its current form is not a good predictor of result alignment.

\textbf{Questionnaire}:
All participants answered the questions about their environment and headphones with ``yes'', while the distribution of answers regarding their experience was almost evenly split (19 ``yes'' versus 14 ``no''). 
Comparing subgroup condition means with respect to their reported experience against the P.800 benchmark showed no statistically significant difference in MAE (two-sided permutation test $p=0.079$).
While these questions are recommended in~\cite{P808},
their suitability as a successful screening method is not confirmed here.

\begin{figure*}[t]
    \centering
    \fontsize{8pt}{9pt}\selectfont 
    \begin{tikzpicture}[every node/.style={font=\footnotesize}]

\definecolor{darkgrey176}{RGB}{176,176,176}
\definecolor{grey}{RGB}{128,128,128}
\definecolor{blue1}{RGB}{34,94,168}
\definecolor{blue2}{RGB}{65,182,196}
\definecolor{blue3}{RGB}{161,158,180}
\definecolor{green1}{RGB}{0,104,55}
\definecolor{green2}{RGB}{49,163,84}
\definecolor{green3}{RGB}{120,198,121}
\definecolor{green4}{RGB}{194,230,153}
\definecolor{magenta1}{RGB}{122,1,119}
\definecolor{magenta2}{RGB}{197,27,138}
\definecolor{magenta3}{RGB}{215,181,216}
\definecolor{magenta4}{RGB}{215,48,31}
\definecolor{magenta5}{RGB}{252,141,89}
\definecolor{magenta6}{RGB}{253,204,138}

\begin{axis}[
tick align=outside,
tick pos=left,
x grid style={darkgrey176},
xmin=0.52, xmax=20.48,
xtick style={color=black},
xtick={1,2,3,4,5,6,7,8,9,10,11,12,13,14,15,16,17,18,19,20},
xticklabel style={rotate=90.0,anchor=east},
xticklabels={
  c01,
  c02,
  c03,
  c04,
  c05,
  c06,
  c07,
  c08,
  c09,
  c10,
  c11,
  c12,
  c13,
  c14,
  c15,
  c16,
  c17,
  c18,
  c19,
  c20,
},
y grid style={darkgrey176},
ylabel={DMOS},
ymajorgrids,
ymin=0.8, ymax=5.45,
ytick style={color=black}
]
\addplot [very thin, grey, dashed]
table {
0.52 0.5
33.48 0.5
};
\addplot [very thin, grey, dashed]
table {
0.52 1.5
33.48 1.5
};
\addplot [very thin, grey, dashed]
table {
0.52 2.5
33.48 2.5
};
\addplot [very thin, grey, dashed]
table {
0.52 3.5
33.48 3.5
};
\addplot [very thin, grey, dashed]
table {
0.52 4.5
33.48 4.5
};
\draw[draw=none,fill=blue1] (axis cs:0.6,0) rectangle (axis cs:1.4,4.48214285714286);
\draw[draw=none,fill=blue2] (axis cs:1.6,0) rectangle (axis cs:2.4,1.83928571428571);
\draw[draw=none,fill=blue2] (axis cs:2.6,0) rectangle (axis cs:3.4,2.80357142857143);
\draw[draw=none,fill=blue2] (axis cs:3.6,0) rectangle (axis cs:4.4,3.94642857142857);
\draw[draw=none,fill=blue3] (axis cs:4.6,0) rectangle (axis cs:5.4,3.16071428571429);
\draw[draw=none,fill=green1] (axis cs:5.6,0) rectangle (axis cs:6.4,2.03571428571429);
\draw[draw=none,fill=green2] (axis cs:6.6,0) rectangle (axis cs:7.4,2.83928571428571);
\draw[draw=none,fill=green2] (axis cs:7.6,0) rectangle (axis cs:8.4,3.07142857142857);
\draw[draw=none,fill=green3] (axis cs:8.6,0) rectangle (axis cs:9.4,3.71428571428571);
\draw[draw=none,fill=green4] (axis cs:9.6,0) rectangle (axis cs:10.4,3.71428571428571);
\draw[draw=none,fill=green4] (axis cs:10.6,0) rectangle (axis cs:11.4,4.28571428571429);
\draw[draw=none,fill=magenta1] (axis cs:11.6,0) rectangle (axis cs:12.4,4.08928571428571);
\draw[draw=none,fill=magenta2] (axis cs:12.6,0) rectangle (axis cs:13.4,3.23214285714286);
\draw[draw=none,fill=magenta3] (axis cs:14.6,0) rectangle (axis cs:15.4,4.14285714285714);
\draw[draw=none,fill=magenta3] (axis cs:13.6,0) rectangle (axis cs:14.4,2.33928571428571);
\draw[draw=none,fill=magenta4] (axis cs:15.6,0) rectangle (axis cs:16.4,3.16071428571429);
\draw[draw=none,fill=magenta4] (axis cs:16.6,0) rectangle (axis cs:17.4,3.98214285714286);
\draw[draw=none,fill=magenta5] (axis cs:17.6,0) rectangle (axis cs:18.4,4.16071428571429);
\draw[draw=none,fill=magenta6] (axis cs:18.6,0) rectangle (axis cs:19.4,3.85714285714286);
\draw[draw=none,fill=magenta6] (axis cs:19.6,0) rectangle (axis cs:20.4,4.10714285714286);
\path [draw=black, semithick]
(axis cs:1,4.35008398597342)
--(axis cs:1,4.61420172831229);

\path [draw=black, semithick]
(axis cs:2,1.59067594339604)
--(axis cs:2,2.08789548517539);

\path [draw=black, semithick]
(axis cs:3,2.51055033097543)
--(axis cs:3,3.09659252616743);

\path [draw=black, semithick]
(axis cs:4,3.74404392414883)
--(axis cs:4,4.14881321870831);

\path [draw=black, semithick]
(axis cs:5,2.84177926810953)
--(axis cs:5,3.47964930331904);

\path [draw=black, semithick]
(axis cs:6,1.75779016264862)
--(axis cs:6,2.31363840877995);

\path [draw=black, semithick]
(axis cs:7,2.54901389809832)
--(axis cs:7,3.12955753047311);

\path [draw=black, semithick]
(axis cs:8,2.75217945240081)
--(axis cs:8,3.39067769045633);

\path [draw=black, semithick]
(axis cs:9,3.44664959768504)
--(axis cs:9,3.98192183088639);

\path [draw=black, semithick]
(axis cs:10,3.4815470781995)
--(axis cs:10,3.94702435037193);

\path [draw=black, semithick]
(axis cs:11,4.13827554650989)
--(axis cs:11,4.43315302491868);

\path [draw=black, semithick]
(axis cs:12,3.90056875450443)
--(axis cs:12,4.278002674067);

\path [draw=black, semithick]
(axis cs:13,2.94526705252921)
--(axis cs:13,3.5190186617565);

\path [draw=black, semithick]
(axis cs:15,3.92844886621907)
--(axis cs:15,4.35726541949522);

\path [draw=black, semithick]
(axis cs:14,1.99225742096709)
--(axis cs:14,2.68631400760434);

\path [draw=black, semithick]
(axis cs:16,2.87477166591412)
--(axis cs:16,3.44665690551445);

\path [draw=black, semithick]
(axis cs:17,3.76164033190479)
--(axis cs:17,4.20264538238092);

\path [draw=black, semithick]
(axis cs:18,3.96872115136734)
--(axis cs:18,4.35270742006123);

\path [draw=black, semithick]
(axis cs:19,3.62594046646632)
--(axis cs:19,4.08834524781939);

\path [draw=black, semithick]
(axis cs:20,3.89128515991029)
--(axis cs:20,4.32300055437542);

\draw (axis cs:1,5.2) ++(0pt,10pt) node[
  scale=1.0,
  anchor=east,
  text=black,
  rotate=90.0
]{4.48};
\draw (axis cs:2,5.2) ++(0pt,10pt) node[
  scale=1.0,
  anchor=east,
  text=black,
  rotate=90.0
]{1.84};
\draw (axis cs:3,5.2) ++(0pt,10pt) node[
  scale=1.0,
  anchor=east,
  text=black,
  rotate=90.0
]{2.80};
\draw (axis cs:4,5.2) ++(0pt,10pt) node[
  scale=1.0,
  anchor=east,
  text=black,
  rotate=90.0
]{3.95};
\draw (axis cs:5,5.2) ++(0pt,10pt) node[
  scale=1.0,
  anchor=east,
  text=black,
  rotate=90.0
]{3.16};
\draw (axis cs:6,5.2) ++(0pt,10pt) node[
  scale=1.0,
  anchor=east,
  text=black,
  rotate=90.0
]{2.04};
\draw (axis cs:7,5.2) ++(0pt,10pt) node[
  scale=1.0,
  anchor=east,
  text=black,
  rotate=90.0
]{2.84};
\draw (axis cs:8,5.2) ++(0pt,10pt) node[
  scale=1.0,
  anchor=east,
  text=black,
  rotate=90.0
]{3.07};
\draw (axis cs:9,5.2) ++(0pt,10pt) node[
  scale=1.0,
  anchor=east,
  text=black,
  rotate=90.0
]{3.71};
\draw (axis cs:10,5.2) ++(0pt,10pt) node[
  scale=1.0,
  anchor=east,
  text=black,
  rotate=90.0
]{3.71};
\draw (axis cs:11,5.2) ++(0pt,10pt) node[
  scale=1.0,
  anchor=east,
  text=black,
  rotate=90.0
]{4.29};
\draw (axis cs:12,5.2) ++(0pt,10pt) node[
  scale=1.0,
  anchor=east,
  text=black,
  rotate=90.0
]{4.09};
\draw (axis cs:13,5.2) ++(0pt,10pt) node[
  scale=1.0,
  anchor=east,
  text=black,
  rotate=90.0
]{3.23};
\draw (axis cs:14,5.2) ++(0pt,10pt) node[
  scale=1.0,
  anchor=east,
  text=black,
  rotate=90.0
]{4.14};
\draw (axis cs:15,5.2) ++(0pt,10pt) node[
  scale=1.0,
  anchor=east,
  text=black,
  rotate=90.0
]{2.34};
\draw (axis cs:16,5.2) ++(0pt,10pt) node[
  scale=1.0,
  anchor=east,
  text=black,
  rotate=90.0
]{3.16};
\draw (axis cs:17,5.2) ++(0pt,10pt) node[
  scale=1.0,
  anchor=east,
  text=black,
  rotate=90.0
]{3.98};
\draw (axis cs:18,5.2) ++(0pt,10pt) node[
  scale=1.0,
  anchor=east,
  text=black,
  rotate=90.0
]{4.16};
\draw (axis cs:19,5.2) ++(0pt,10pt) node[
  scale=1.0,
  anchor=east,
  text=black,
  rotate=90.0
]{3.86};
\draw (axis cs:20,5.2) ++(0pt,10pt) node[
  scale=1.0,
  anchor=east,
  text=black,
  rotate=90.0
]{4.11};
\end{axis}

\end{tikzpicture}
    \hspace{0.9cm}
    \begin{tikzpicture}[every node/.style={font=\footnotesize}]

\definecolor{darkgrey176}{RGB}{176,176,176}
\definecolor{grey}{RGB}{128,128,128}
\definecolor{blue1}{RGB}{34,94,168}
\definecolor{blue2}{RGB}{65,182,196}
\definecolor{blue3}{RGB}{161,158,180}
\definecolor{green1}{RGB}{0,104,55}
\definecolor{green2}{RGB}{49,163,84}
\definecolor{green3}{RGB}{120,198,121}
\definecolor{green4}{RGB}{194,230,153}
\definecolor{magenta1}{RGB}{122,1,119}
\definecolor{magenta2}{RGB}{197,27,138}
\definecolor{magenta3}{RGB}{215,181,216}
\definecolor{magenta4}{RGB}{215,48,31}
\definecolor{magenta5}{RGB}{252,141,89}
\definecolor{magenta6}{RGB}{253,204,138}

\begin{axis}[
tick align=outside,
tick pos=left,
x grid style={darkgrey176},
xmin=0.52, xmax=20.48,
xtick style={color=black},
xtick={1,2,3,4,5,6,7,8,9,10,11,12,13,14,15,16,17,18,19,20},
xticklabel style={rotate=90.0,anchor=east},
xticklabels={
  c01,
  c02,
  c03,
  c04,
  c05,
  c06,
  c07,
  c08,
  c09,
  c10,
  c11,
  c12,
  c13,
  c14,
  c15,
  c16,
  c17,
  c18,
  c19,
  c20,
},
y grid style={darkgrey176},
ylabel={DMOS},
ymajorgrids,
ymin=0.8, ymax=5.45,
ytick style={color=black}
]
\addplot [very thin, grey, dashed]
table {
0.52 0.5
33.48 0.5
};
\addplot [very thin, grey, dashed]
table {
0.52 1.5
33.48 1.5
};
\addplot [very thin, grey, dashed]
table {
0.52 2.5
33.48 2.5
};
\addplot [very thin, grey, dashed]
table {
0.52 3.5
33.48 3.5
};
\addplot [very thin, grey, dashed]
table {
0.52 4.5
33.48 4.5
};
\draw[draw=none,fill=blue1] (axis cs:0.6,0) rectangle (axis cs:1.4,4.71428571428571);
\draw[draw=none,fill=blue2] (axis cs:1.6,0) rectangle (axis cs:2.4,1.17857142857143);
\draw[draw=none,fill=blue2] (axis cs:2.6,0) rectangle (axis cs:3.4,2.28571428571429);
\draw[draw=none,fill=blue2] (axis cs:3.6,0) rectangle (axis cs:4.4,3.71428571428571);
\draw[draw=none,fill=blue3] (axis cs:4.6,0) rectangle (axis cs:5.4,2.71428571428571);
\draw[draw=none,fill=green1] (axis cs:5.6,0) rectangle (axis cs:6.4,1.46428571428571);
\draw[draw=none,fill=green2] (axis cs:6.6,0) rectangle (axis cs:7.4,2.78571428571429);
\draw[draw=none,fill=green2] (axis cs:7.6,0) rectangle (axis cs:8.4,2.92857142857143);
\draw[draw=none,fill=green3] (axis cs:8.6,0) rectangle (axis cs:9.4,3.5);
\draw[draw=none,fill=green4] (axis cs:9.6,0) rectangle (axis cs:10.4,3.60714285714286);
\draw[draw=none,fill=green4] (axis cs:10.6,0) rectangle (axis cs:11.4,4.39285714285714);
\draw[draw=none,fill=magenta1] (axis cs:11.6,0) rectangle (axis cs:12.4,4.17857142857143);
\draw[draw=none,fill=magenta2] (axis cs:12.6,0) rectangle (axis cs:13.4,2.92857142857143);
\draw[draw=none,fill=magenta3] (axis cs:14.6,0) rectangle (axis cs:15.4,3.92857142857143);
\draw[draw=none,fill=magenta3] (axis cs:13.6,0) rectangle (axis cs:14.4,1.64285714285714);
\draw[draw=none,fill=magenta4] (axis cs:15.6,0) rectangle (axis cs:16.4,2.78571428571429);
\draw[draw=none,fill=magenta4] (axis cs:16.6,0) rectangle (axis cs:17.4,3.96428571428571);
\draw[draw=none,fill=magenta5] (axis cs:17.6,0) rectangle (axis cs:18.4,4.10714285714286);
\draw[draw=none,fill=magenta6] (axis cs:18.6,0) rectangle (axis cs:19.4,3.67857142857143);
\draw[draw=none,fill=magenta6] (axis cs:19.6,0) rectangle (axis cs:20.4,4.21428571428571);
\path [draw=black, semithick]
(axis cs:1,4.54388314083966)
--(axis cs:1,4.88468828773177);

\path [draw=black, semithick]
(axis cs:2,1.03410561818582)
--(axis cs:2,1.32303723895704);

\path [draw=black, semithick]
(axis cs:3,1.88427243991896)
--(axis cs:3,2.68715613150961);

\path [draw=black, semithick]
(axis cs:4,3.43170501456995)
--(axis cs:4,3.99686641400148);

\path [draw=black, semithick]
(axis cs:5,2.31284386849039)
--(axis cs:5,3.11572756008104);

\path [draw=black, semithick]
(axis cs:6,1.17068104180296)
--(axis cs:6,1.75789038676847);

\path [draw=black, semithick]
(axis cs:7,2.35378493621027)
--(axis cs:7,3.2176436352183);

\path [draw=black, semithick]
(axis cs:8,2.51378902243003)
--(axis cs:8,3.34335383471282);

\path [draw=black, semithick]
(axis cs:9,3.12959481645096)
--(axis cs:9,3.87040518354904);

\path [draw=black, semithick]
(axis cs:10,3.28298583072236)
--(axis cs:10,3.93129988356335);

\path [draw=black, semithick]
(axis cs:11,4.20863706513206)
--(axis cs:11,4.57707722058223);

\path [draw=black, semithick]
(axis cs:12,3.93053469294095)
--(axis cs:12,4.4266081642019);

\path [draw=black, semithick]
(axis cs:13,2.52622645037302)
--(axis cs:13,3.33091640676984);

\path [draw=black, semithick]
(axis cs:15,3.61091799524557)
--(axis cs:15,4.24622486189728);

\path [draw=black, semithick]
(axis cs:14,1.23691972750204)
--(axis cs:14,2.04879455821225);

\path [draw=black, semithick]
(axis cs:16,2.36571428571429)
--(axis cs:16,3.20571428571429);

\path [draw=black, semithick]
(axis cs:17,3.63789694360016)
--(axis cs:17,4.29067448497127);

\path [draw=black, semithick]
(axis cs:18,3.81602114401954)
--(axis cs:18,4.39826457026617);

\path [draw=black, semithick]
(axis cs:19,3.31434215604798)
--(axis cs:19,4.04280070109488);

\path [draw=black, semithick]
(axis cs:20,3.8898488802806)
--(axis cs:20,4.53872254829082);

\draw (axis cs:1,5.2) ++(0pt,10pt) node[
  scale=1.0,
  anchor=east,
  text=black,
  rotate=90.0
]{4.71};
\draw (axis cs:2,5.2) ++(0pt,10pt) node[
  scale=1.0,
  anchor=east,
  text=black,
  rotate=90.0
]{1.18};
\draw (axis cs:3,5.2) ++(0pt,10pt) node[
  scale=1.0,
  anchor=east,
  text=black,
  rotate=90.0
]{2.29};
\draw (axis cs:4,5.2) ++(0pt,10pt) node[
  scale=1.0,
  anchor=east,
  text=black,
  rotate=90.0
]{3.71};
\draw (axis cs:5,5.2) ++(0pt,10pt) node[
  scale=1.0,
  anchor=east,
  text=black,
  rotate=90.0
]{2.71};
\draw (axis cs:6,5.2) ++(0pt,10pt) node[
  scale=1.0,
  anchor=east,
  text=black,
  rotate=90.0
]{1.46};
\draw (axis cs:7,5.2) ++(0pt,10pt) node[
  scale=1.0,
  anchor=east,
  text=black,
  rotate=90.0
]{2.79};
\draw (axis cs:8,5.2) ++(0pt,10pt) node[
  scale=1.0,
  anchor=east,
  text=black,
  rotate=90.0
]{2.93};
\draw (axis cs:9,5.2) ++(0pt,10pt) node[
  scale=1.0,
  anchor=east,
  text=black,
  rotate=90.0
]{3.50};
\draw (axis cs:10,5.2) ++(0pt,10pt) node[
  scale=1.0,
  anchor=east,
  text=black,
  rotate=90.0
]{3.61};
\draw (axis cs:11,5.2) ++(0pt,10pt) node[
  scale=1.0,
  anchor=east,
  text=black,
  rotate=90.0
]{4.39};
\draw (axis cs:12,5.2) ++(0pt,10pt) node[
  scale=1.0,
  anchor=east,
  text=black,
  rotate=90.0
]{4.18};
\draw (axis cs:13,5.2) ++(0pt,10pt) node[
  scale=1.0,
  anchor=east,
  text=black,
  rotate=90.0
]{2.93};
\draw (axis cs:14,5.2) ++(0pt,10pt) node[
  scale=1.0,
  anchor=east,
  text=black,
  rotate=90.0
]{3.93};
\draw (axis cs:15,5.2) ++(0pt,10pt) node[
  scale=1.0,
  anchor=east,
  text=black,
  rotate=90.0
]{1.64};
\draw (axis cs:16,5.2) ++(0pt,10pt) node[
  scale=1.0,
  anchor=east,
  text=black,
  rotate=90.0
]{2.79};
\draw (axis cs:17,5.2) ++(0pt,10pt) node[
  scale=1.0,
  anchor=east,
  text=black,
  rotate=90.0
]{3.96};
\draw (axis cs:18,5.2) ++(0pt,10pt) node[
  scale=1.0,
  anchor=east,
  text=black,
  rotate=90.0
]{4.11};
\draw (axis cs:19,5.2) ++(0pt,10pt) node[
  scale=1.0,
  anchor=east,
  text=black,
  rotate=90.0
]{3.68};
\draw (axis cs:20,5.2) ++(0pt,10pt) node[
  scale=1.0,
  anchor=east,
  text=black,
  rotate=90.0
]{4.21};
\end{axis}

\end{tikzpicture}
    \caption{P.808 results after mid- (left) and post-screening (right).}
    \label{fig:afterscreen}
\end{figure*}
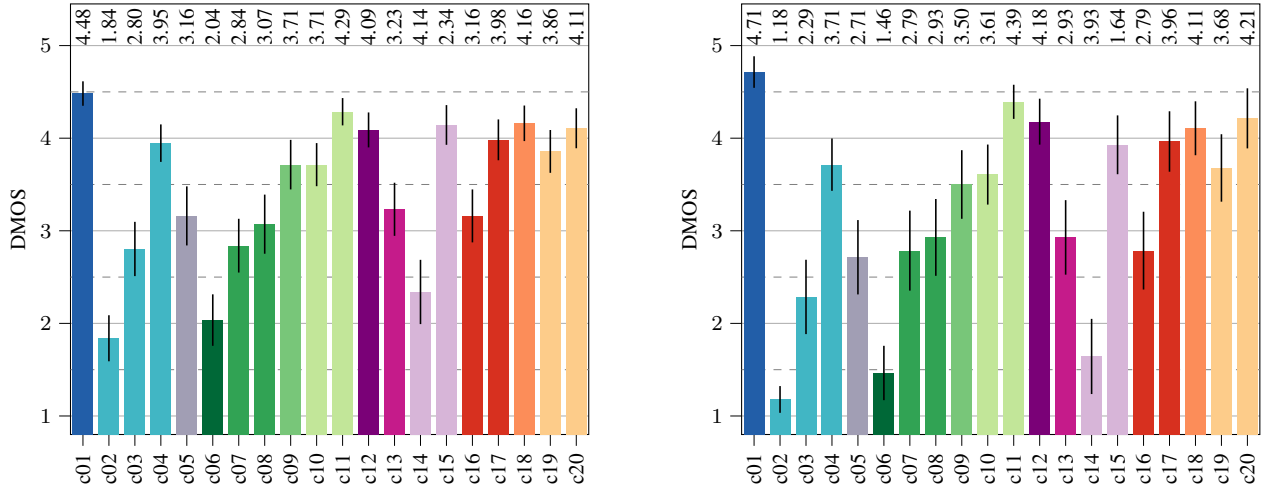

\subsection{Analysis of mid-screening methods}
\textbf{Traps}:
Most of the traps were answered correctly, which confirms the participants' attention. 
Nevertheless, we were able to identify one extreme outlier that failed the majority of traps.\footnote{The reason might be a lack of English skill and a misinterpretation of the scale since all actual results seemed inverted.} 
This can be a reliable tool for detecting strong outliers. 
An in-depth evaluation of this screening method is found in~\cite{naderi15_interspeech}.

\textbf{Gold standard question}:
The mean of the lowest reference ratings between the two tests differ significantly: $4.44$ and $2.97$ DMOS points for the P.800 and P.808 test, respectively.
When screening based on this parameter, all measures improve monotonically when increasing the minimum reference threshold. 
A threshold of $4$ improves the values to $\mathrm{MAE}=0.327$, $\mathrm{RMSE}=0.401$, $r=0.963$, $\rho=0.963$, while retaining 14 participants.
This screening is highly effective and satisfies its goal to decrease the variance for all conditions by enforcing a reduced spread of the reference ratings.

\subsection{Analysis of post-screening methods}
\textbf{Rating span}:
In the laboratory test, the rating span ranges from $2.25$ to $4$, while the crowdsourced test yields spans starting at $-2.5$, indicating a reversal of perceived high and low quality. 
In total, four participants obtained negative rating spans and only one participant achieved the ideal value of $4$. 
When screening based on this method, all measures improve monotonically for increasingly strict thresholds up to a rating span of $3.25$ (up to some slight deviations due to noise).
Enforcing higher rating spans excludes too many participants to be able to draw meaningful conclusions. 
It would anyway not be beneficial to enforce the theoretically ideal span of $4$ since anchor ratings are highly subjective. 
For a threshold of $2.5$, the quality already improves significantly to $\mathrm{MAE}=0.284$, $\mathrm{RMSE}=0.325$, $r=0.956$, $\rho=0.962$, while retaining a reasonable amount of ten participants. 
This means the rating span screening can substantially improve the alignment with the P.800 test and serves its purpose of increasing the resolution of condition ratings.

\textbf{Anchor ordering}:
In the P.800 listening test, all 27 participants attained a perfect anchor ordering score, while only 15 out of the 33 P.808 participants managed to achieve that.
Applying increasingly strict anchor ordering thresholds yields increasing improvements in agreement with the laboratory test.
Screening for perfect ordering leads to a significant enhancement of the results: $\text{MAE}=0.376$, $\text{RMSE}=0.428$, $r=0.962$, $\rho=0.942$.
From a practical point of view, we recommend using an MNRU ranking threshold greater than or equal to the number of MNRU conditions minus one. 
In our test, a threshold of $2$ 
yields a meaningful reduction in error while retaining most participants.

\subsection{Combined results}
While the pre-screening results fall short of expectations, the mid-screening and post-screening results are very effective. Figure~\ref{fig:afterscreen} shows the results after mid-screening with the gold standard reference minimum rating of $4$ and the exclusion of the trap outlier on the left. Note that the trap outlier also failed the gold standard questions.
This retains 14 participants.
The results after post-screening are shown on the right side for a rating span threshold of 2.5 and a perfect MNRU ordering.
This retains seven participants while improving the measures to $\mathrm{MAE}=0.230$, $\mathrm{RMSE}=0.259$, $r=0.974$, and $\rho =0.958$.
This screening only retains 7 participants.
It is clearly visible that both screening categories significantly improve the quality of the test results, while the post-screening appears even more powerful at the cost of further reducing the number of remaining participants. 
To still retain reliable results for all conditions, it is necessary to recruit three to five times the number of desired participants as recommended in~\cite{P808}.
All of the remaining participants after post-screening also passed the mid-screening which means that mid-screening can be a good indicator to detect well-performing participants early.
There is some overlap between the participants screened by the two proposed post-screening methods, but one is not a subset of the other. 
The two methods complement each other as one enforces use of the (almost) full scale while the other verifies the correct ordering of conditions.
This intuition is supported by the improvement of MAE, RMSE, and $r$ correlation values for the combined post-screening compared to the individual screening methods, while keeping a high $\rho$ value.
P.808 recommends to only screen for outliers based on standardized outlier rules. 
However, for high agreement among participants within one condition, the within-condition standard deviation is very small.
As a result, standardized outlier rules can mistakenly flag moderate deviations as outliers even if those deviations reflect genuine perceptual differences between raters.
Instead, we recommend using these four screening methods with predefined thresholds to make user-based screening of P.808 results easier, automatic, bias-free, and uniform.

\section{Conclusion}

In this paper, we highlighted that drawbacks from using crowdsourcing platforms for evaluating speech and audio codecs can be compensated by incorporating suitable screening methods.
We showed that the tested pre-screening has no effect, while mid- and post-screening directly correlates with the quality of the P.808 results.
Consequently, we proposed to employ screening in the form of traps, minimum reference rating, rating scale, and anchor ordering to better align the crowdsourced results with P.800 test results.
In doing so, crowdsourced listening tests become more reliable and viable as a substitution for the more costly laboratory tests, especially for early prototype evaluations.
To build upon the insights gained, future work may include investigations of adjusted questionnaires and pretests, different languages, general audio, and stereo material.
\vfill\pagebreak
\section{Acknowledgments}

\ifcameraready
     This work has been supported by the Free State of Bavaria in the DSgenAI project.
     The authors thank Nathan Cormier for contributing to this work during his internship at Fraunhofer IIS.
     The authors further thank Maximilian Schlegel, Guillaume Fuchs, and Markus Multrus for their invaluable support and feedback.
\else
     The authors thank several people.
\fi

\section{Generative AI Use Disclosure}
No Generative AI was used in the preparation of this paper.

\flushend
\bibliographystyle{IEEEtran}
\bibliography{mybib}

\end{document}